\documentstyle[11pt,aaspp4,flushrt]{article}
\lefthead{Taylor, Thorstensen, \& Patterson}
\righthead{ LS Peg: A Low-Inclination SW Sextantis-Type Cataclysmic Binary
with High-Velocity Balmer Emission Line Wings} 
\slugcomment{To appear in: P.A.S.P., February 1999}

\begin{document}

\title{LS Peg: A Low-Inclination SW Sextantis-Type Cataclysmic Binary with
High-Velocity Balmer Emission Line Wings \footnote
{Based in part on observations obtained at MDM Observatory, operated by  
Dartmouth College, Columbia University, the Ohio State University, and the
University of Michigan.}}
\author{Cynthia J. Taylor, John R. Thorstensen,} 
\affil{Department of Physics and Astronomy, Dartmouth College, 
6127 Wilder Laboratories\\
 Hanover, NH 03755-3528; {\tt cynthia.j.taylor, thorstensen@dartmouth.edu}}
\author{and Joseph Patterson}
\affil{Department of Astronomy,
Columbia University, 550 West 120th Street\\
New York, NY 10027;
{\tt jop@astro.columbia.edu}}

\begin{abstract}
We present time-resolved spectroscopy and photometry of the bright cataclysmic
variable LS Peg (= S193; V $\approx$ 13.0 -- Szkody et al. 
\markcite{szg97}1997).  The Balmer lines exhibit broad, asymmetric wings
Doppler-shifted by about 2000 km s$^{-1}$ at the edges, while the \ion{He}{1} 
lines show phase-dependent absorption features strikingly similar to SW 
Sextantis stars, as well as emission through most of the phase.  The 
\ion{C}{3}/\ion{N}{3} emission blend
does not show any phase dependence.  From velocities of H$\alpha$ emission 
lines, we determine an orbital period of $0.174774 \pm 0.000003$ d (= 4.1946 
h), which agrees with Szkody's \markcite{sz95}(1995) value of approximately 
4.2 hours. No stable photometric signal was found at the orbital period. 
A non-coherent quasi-periodic photometric signal was seen at a period of 
$20.7 \pm 0.3$ min.

The high-velocity Balmer wings most probably arise from a stream re-impact 
point close to the white dwarf.  We present simulated spectra
based on a kinematic model similar to the modified 
disk-overflow scenario of Hellier \& Robinson \markcite{hr94}(1994).  The
models reproduce the broad line wings, though some other details are 
unexplained.

Using an estimate of dynamical phase based on the model, we show that the
phasing of the emission- and absorption-line variations is consistent with
that in (eclipsing) SW Sex stars.  We therefore identify LS Peg as a 
low-inclination SW Sex star.

Our model suggests $i = 30\arcdeg$, and the observed absence of any 
photometric signal at the orbital frequency establishes $i < 60\arcdeg$. This
constraint puts a severe strain on interpretations of the SW Sex phenomenon
which rely on disk structures lying slightly out of the orbital plane.

\end{abstract}

\section{Introduction}

LS Peg (S193) was first identified as a cataclysmic variable (CV) by Downes \& 
Keyes \markcite{dk88}(1988). It is among the brightest CVs in the sky (V
$\approx$ 13.0 -- Szkody et al. \markcite{szg97}1997) which lacks
a well-established orbital period.  It is apparently a novalike 
system; such systems often show relatively high excitation emission lines,
long-term brightness variations, and, in the case of the SW Sex stars, 
orbital variations in the narrow absorption components of \ion{He}{1} and
radial velocity phase shifts with respect to the photometric eclipse phase
(Warner \markcite{w95}1995; Thorstensen et al. \markcite{tr91}1991).

The classification of SW Sextantis (SW Sex) stars was created by Thorstensen
et al. \markcite{tr91}(1991). This classification has been somewhat
controversial because it is based on spectroscopic properties while other
variable-star classifications are based on photometric properties;
therefore the SW Sex classification overlaps with other classifications
(Dhillon \markcite{dhil96}1996).
SW Sex stars were originally defined as 
novalike variables with 1) eclipses, 2) orbital periods between 3 and 4 hours, 
3) \ion{He}{2} $\lambda$4686 emission around half the strength of H$\beta$, 
4) a Balmer-line radial velocity phase that lags substantially behind 
the orbital phase, 5) absorption in Balmer and \ion{He}{1} lines opposite
the eclipse. This last characteristic is known as ``phase 0.5 absorption'',
referring to eclipse phase as phase $\phi = 0.$ Over time, the definition has
simplified somewhat, to: novalikes, with phase-dependent absorption around 
phase 0.5, in which the spectroscopic phase lags significantly behind the 
orbital phase (Warner \markcite{w95}1995, Dhillon \markcite{dhil96}1996). 
Since inclination depends on line of sight from Earth, the ``eclipsing'' 
characteristic has increasingly fallen away. In addition to the defining
characteristics, SW Sex stars tend to have continuum and/or high-excitation
lines eclipsed more strongly than low excitation lines (Warner
\markcite{w95}1995 and references therein; Young, Schneider, \& Shectman
\markcite{yss81}1981;  Baptista, Steiner, \& Horne \markcite{bsh96}1996),
relatively narrow FWHM ($\sim 1000$ km s$^{-1}$) and large FWZI in the Balmer
lines, no observable circular polarization (Young, Schneider, \& Shectman
\markcite{yss81}1981; Grauer et al. \markcite{grwlsg94}1994; Thorstensen et al.
\markcite{tr91}1991; Dhillon \& Rutten \markcite{dr95}1995; Stockman et al.
\markcite{st92}1992), and a radial temperature profile in the inner disk
which is noticably flatter than $T \propto R^{-3/4}$ while in the high state
(Baptista, Steiner, \& Horne \markcite{bsh96}1996; Rutten, van Paradijs, \&
Tinbergen \markcite{rvt92}1992). 

There is no single comprehensive study of LS Peg in the literature, but many
observations have been published and we summarize these here.  No stable 
photometric periods are known, but plenty of quasi-periodic oscillations 
have been seen in the light curve at different periods: a $16.5 \pm 2$ min 
period from a one-night optical light curve, an intermittent 18.7 min period 
over several nights, a weaker quasi period at 32 min, and other possible 
periods ranging from 5 to 50 min (Garnavich, Szkody, \& Goldader 
\markcite{gsg88}1988; Garnavich \& Szkody \markcite{gs92}1992; Szkody et al. 
\markcite{sz94}1994a).  
No significant circular polarization was detected (Stockman et al. 
\markcite{st92}1992).  Analysis of the Harvard plate collections from 
1897 -- 1988 showed long term high and low states of B $\approx$ 12 and B 
$\approx$ 14 (Garnavich \& Szkody \markcite{gs92}1992).  
{\it IUE} and {\it Voyager} spectra of LS Peg in a high state were similar to 
SW Sex candidates V795 Her and PG0859+415 (BP Lyn), with all these showing
deep UV absorption lines. The line ratios and slope of the UV flux 
distribution are not typical of dwarf novae at outburst. The low-state spectrum
is most similar to 1H0551-819, with a flat flux distribution, only \ion{C}{4} 
and \ion{Mg}{2} lines in emission, and \ion{Si}{3}
and \ion{N}{5} in deep absorption (Szkody \markcite{sz91}1991; Szkody et 
al. \markcite{szg97}1997). No X-ray period was determined from {\it Ginga}
and {\it ROSAT} data, but the data place an upper limit of 40\% on the 
sinusoidal amplitude.  The differences in column densities and temperatures 
between the {\it ROSAT} and {\it Ginga} data indicated either a two-component 
system or a change of accretion characteristics (Szkody et 
al. \markcite{sz94}1994a;  Szkody et al. \markcite{sgcf94}1994b).  
Szkody \markcite{sz91}(1991) reported that the velocities of the Balmer-line 
wings $< 50$ km s$^{-1}$, which implies a low inclination. In addition, very
prominent, highly variable (on timescales $< 10$ minutes), and random
absorptions in \ion{He}{1} $\lambda$4471 and Balmer emission were seen, with
the absorption always redshifted by 250-300 km s$^{-1}$. By 1995, Balmer-line 
wings had appeared showing an extremely large velocity amplitude of $\sim$1700 
km s$^{-1}$; these were the only sinusoidal features seen (Szkody 
\markcite{sz95}1995). No orbital period has been formally published, but 
Szkody et al. \markcite{sgcf94}(1994b) and Szkody \markcite{sz95}(1995) refer
to an orbital period of $\sim$4.2 hours from the Balmer emission-line wings.

\section{Observations and Reductions}
\subsection{Spectroscopy}

We used the telescopes at MDM Observatory on Kitt 
Peak, during five observing runs -- 1996 July, 1996 September, 1996 October, 
1996 December, and 1997 June.  Table~\protect\ref{tbl-1} lists the different 
combinations of telescopes, spectrographs, and CCD chips, as well as the 
spectral resolution (FWHM). The exposure times ranged from 5 to 8 minutes.  
The exact dates of observation are implicit in the radial velocity timeseries 
(Table~\protect\ref{tbl-2}).  To avoid daily cycle count ambiguity, we 
observed over a wide range of hour angle.  We derived the wavelength scale  
from comparison lamp spectra bracketing the observations as the telescope 
tracked. Flux standards were observed when conditions warrented.

For data reduction we used standard IRAF routines\footnote{IRAF is 
distributed by the National Optical Astronomy Observatories.}.  The 
pixel-to-wavelength transformations had typical RMS residuals of $< 0.1$ \AA.  
The stability of the 5577.350 \AA\ night-sky line in the reduced data was
$< 5$ km s$^{-1}$, and the systematic uncertainty was on the order of 10 km 
s$^{-1}$. We estimate our flux calibrations to be accurate to about 30 per cent.

Because most of the lines have an asymmetric wing component, we could not use 
the standard algorithms described by Schneider \& Young \markcite{sy80}(1980) 
to measure the radial velocities.  Since the base of the lines have an 
approximately constant width (see Figure~\protect\ref{fig6}) and the wings 
shift back and forth, the radial velocities were measured by hand using the 
midpoint of the full width zero intensity (FWZI).  The spectra were first 
rectified using the IRAF task {\it sfit} and the FWZI were measured 
in {\it splot}. The times and velocities were corrected to the solar center.

\begin{figure}
\plotfiddle{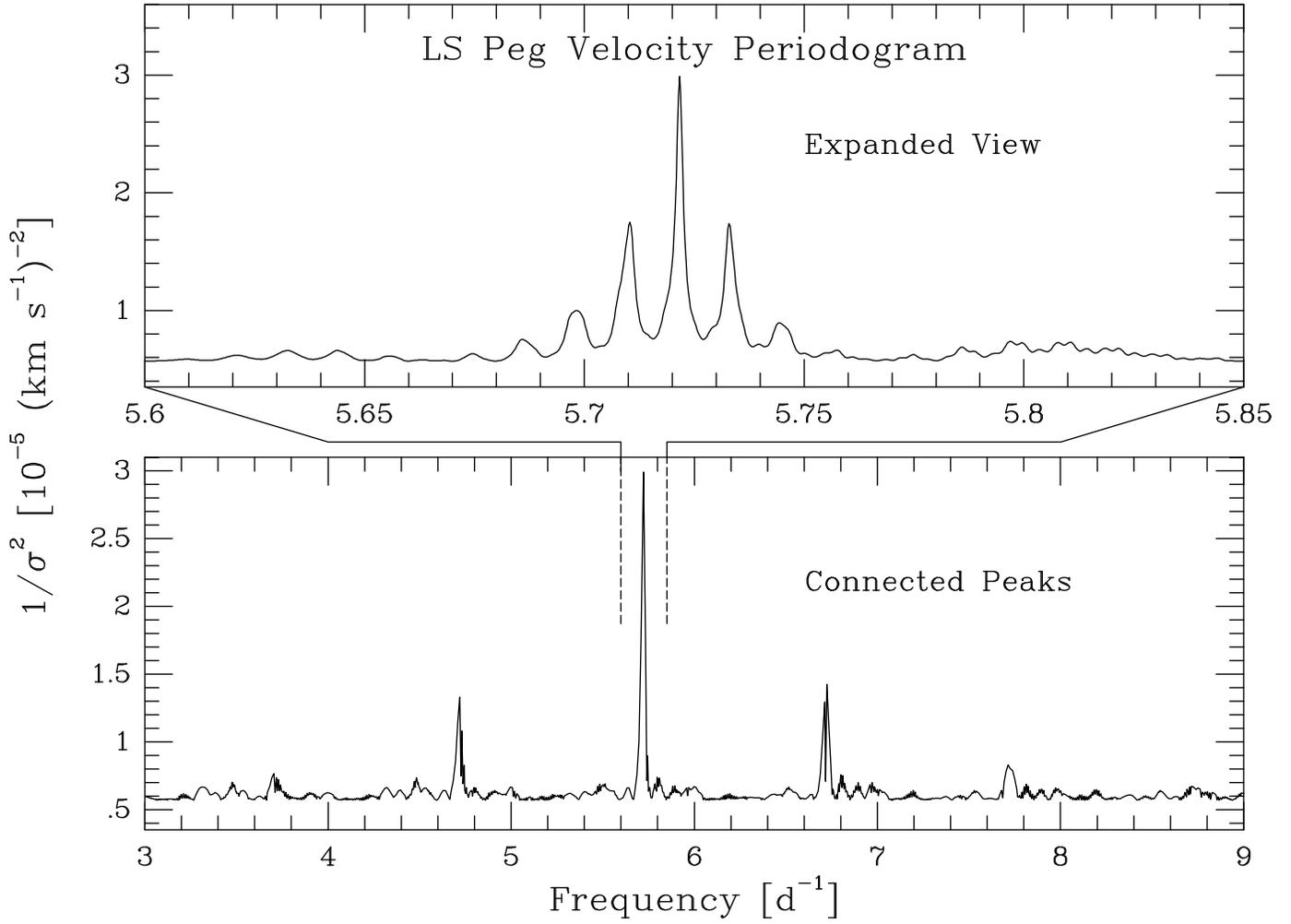}{4.0truein}{270}{70}{70}{-270}{420}
\caption{Periodogram of velocities of LS Peg.  The lower panel shows the
upper envelope of the periodogram, created by joining local maxima with 
straight lines, while the upper shows a magnified view of the region of the 
highest peak. \label{fig1}}
\end{figure}

Figure~\protect\ref{fig1} shows results of a period search conducted using the
`residual-gram' method described by Thorstensen et 
al. \markcite{t96}(1996).   The best fit is at 5.722 cycles/day. The Monte 
Carlo test of Thorstensen \& Freed \markcite{tf85}(1985) shows that the cycle 
count is unambiguous on all scales.  A least-squares sinusoid fit of 
\begin{displaymath}
v(t)=\gamma +K\sin[2\pi (t-T_{\rm 0})/P]
\end{displaymath}
yielded
\begin{displaymath}
\begin{array}{rcl}
T_0 & = & \hbox{HJD}\ 2450358.8964 \pm 0.0009, \\
P & = & 0.174774 \pm 0.000003\ {\rm d}, \\
K  & = & 593 \pm 21\ {\rm km\ s^{-1}}, \\
\gamma & = & -56 \pm 14\ {\rm km\ s^{-1}}, \ \hbox{and} \\
\sigma & = & 183\ {\rm km\ s^{-1}},
\end{array}
\end{displaymath}
where $\sigma$ is the uncertainty of a single measurement inferred
from the goodness-of-fit. 

\begin{figure}
\plotfiddle{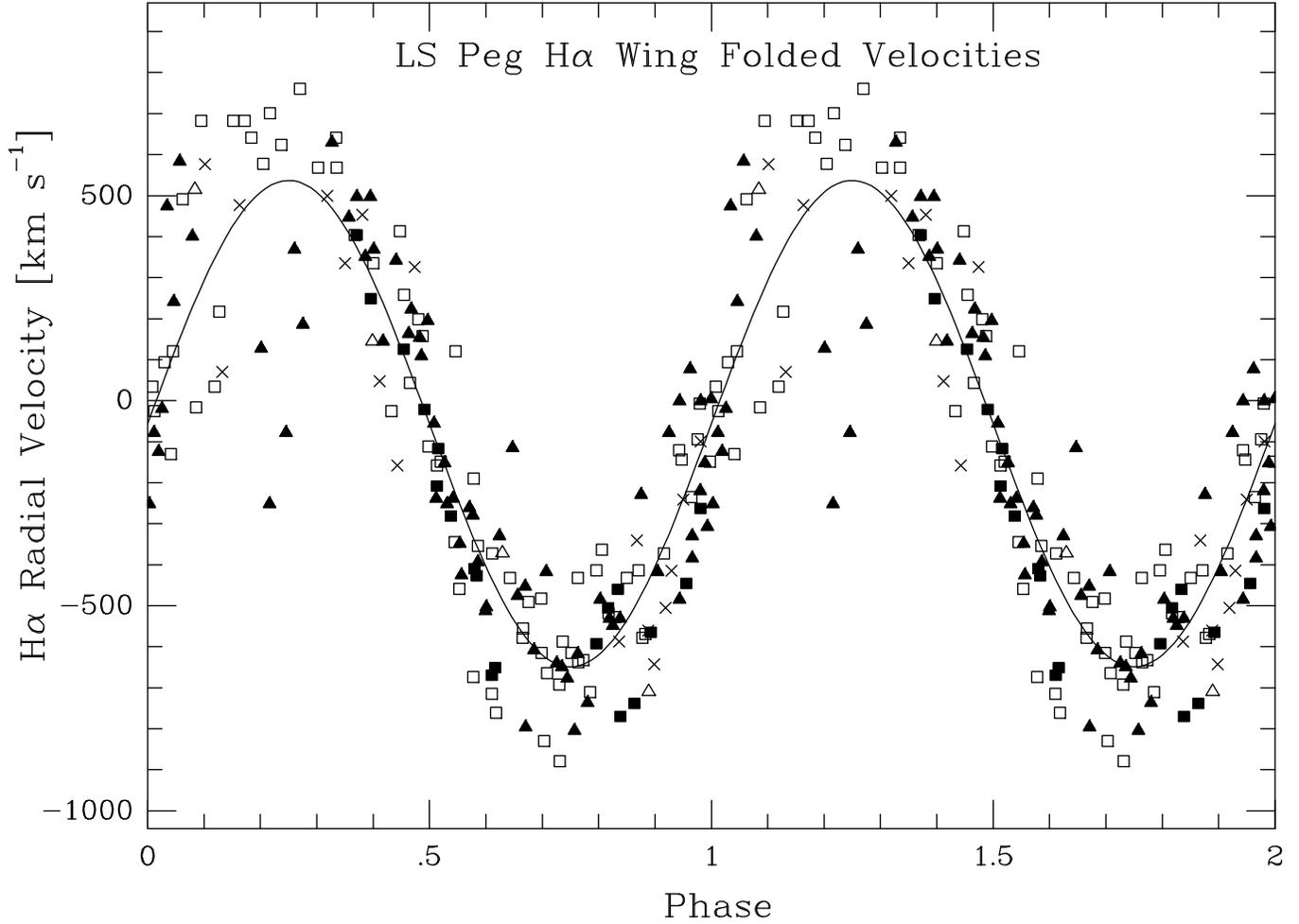}{4.0truein}{270}{70}{70}{-270}{420}
\figcaption{H$\alpha$ radial velocities derived from the broad emission wings 
folded on the ephemerides given in text.  Two cycles are plotted for continuity,
and the best--fit sinusoid is shown.  The symbols are: solid squares, 1996 
July; squares, 1996 September; solid triangles, 1996 October; crosses, 1996 
December; triangles, 1997 June. \label{fig2}}
\end{figure}

Figure~\protect\ref{fig2} shows the velocities derived from the H$\alpha$ wings
folded on the best-fit period with the sinusoidal fit superposed.  The  $K$ 
value is far too large for any plausible motion of the white dwarf.
Interpreted literally, this $K$ value yields a secondary
mass function of $f(M_2) = 3.77 M_{\sun}$, which, as an illustration, gives 
$M_2 \gtrsim 30 M_{\sun}$ assuming $i = 30\arcdeg$ and $M_1 = 0.7 M_{\sun}$.

\subsection{Photometry}
\begin{figure}
\plotfiddle{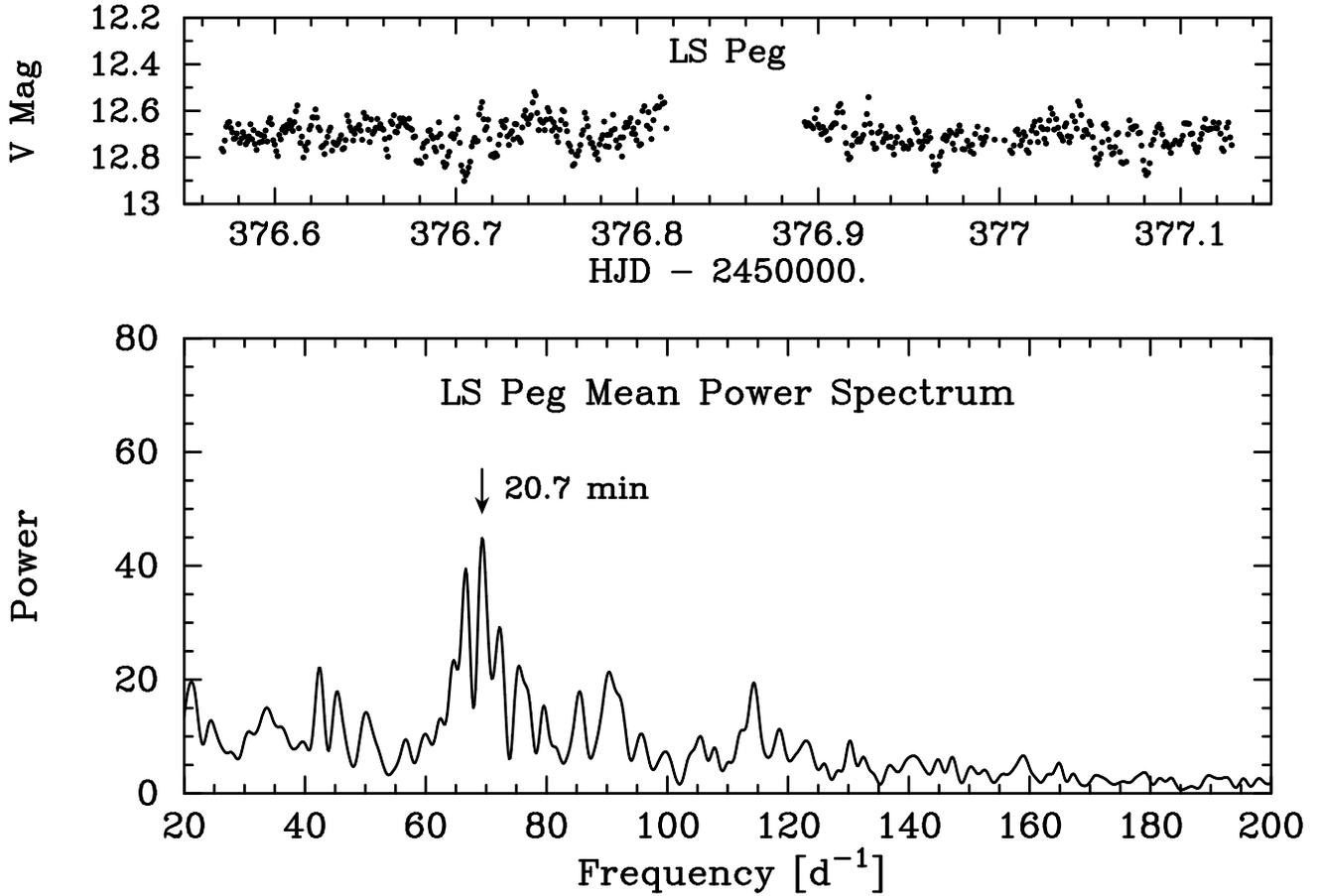}{4.0truein}{270}{70}{70}{-270}{420}
\caption{Upper frame: sample light curve of LS Peg.  Lower frame: mean
power spectrum of LS Peg from the 6 longest nights of photometry, with the 
main ``QPO'' signal marked with its period in minutes.  The power in this
signal corresponds to a full amplitude of 0.05 mag, but this must underestimate
the actual amplitude, since the frequency rapidly wanders.  \label{fig3}}
\end{figure}

 We carried out time-series photometry of LS Peg for 12 nights
(distibuted over an interval of 18 nights) in 1996 October, using the
telescopes of the Center for Backyard Astrophysics (Harvey et 
al. \markcite{h95}1995).  We accumulated 55 hr of coverage, during which the 
star varied irregularly in the range V=12.5-13. A sample light curve is shown
in the top frame of Figure~\protect\ref{fig3}.

We searched for periodic signals by calculating the power spectrum (from the
discrete Fourier transform), and found no stable periods.  The full
amplitude upper limit to any signal in the vicinity of $P_{\rm orb}$ was 0.04
mag, and the upper limit at $P_{\rm orb}$ exactly was 0.027 mag.

     The individual nights showed features in the power spectrum near 70 cycles
d$^{-1}$, and a weighted average over the 6 longest nights of coverage
yielded the mean power spectrum in the lower frame of 
Figure~\protect\ref{fig3} .  The highest peak occurs at a period of $20.7 \pm 
0.3$ min, and corresponds to a full amplitude of 0.041 mag in the 12-night
time series.  However, the structure and precise location of the 
peak are not stable, but drift about on a timescale of hours.  Thus this is 
probably a manifestation of the signal reported at 19 min by Garnavich \& 
Szkody \markcite{gs92}(1992), with the period difference merely arising from 
erratic drifts in the star.  Since the signal is not stable in phase, the
measured amplitude of 0.041 mag underestimates the true amplitude of the
(incoherent) signal, which occasionally appears to reach 0.2 mag in the raw
light curve.

In CV parlance, such things are called quasi-periodic oscillations
(QPOs).  They are commonly seen in novalike variables (TT Ari: Andronov
et al. \markcite{a98}1998; V795 Her: Zhang et al. \markcite{z91}1991; 
AH Men: Patterson \markcite{p95}1995), and are oddly
clustered near 20 min. But there is still no understanding of the
origin of these signals.

\section{Analysis of Spectra}

\begin{figure}
\plotfiddle{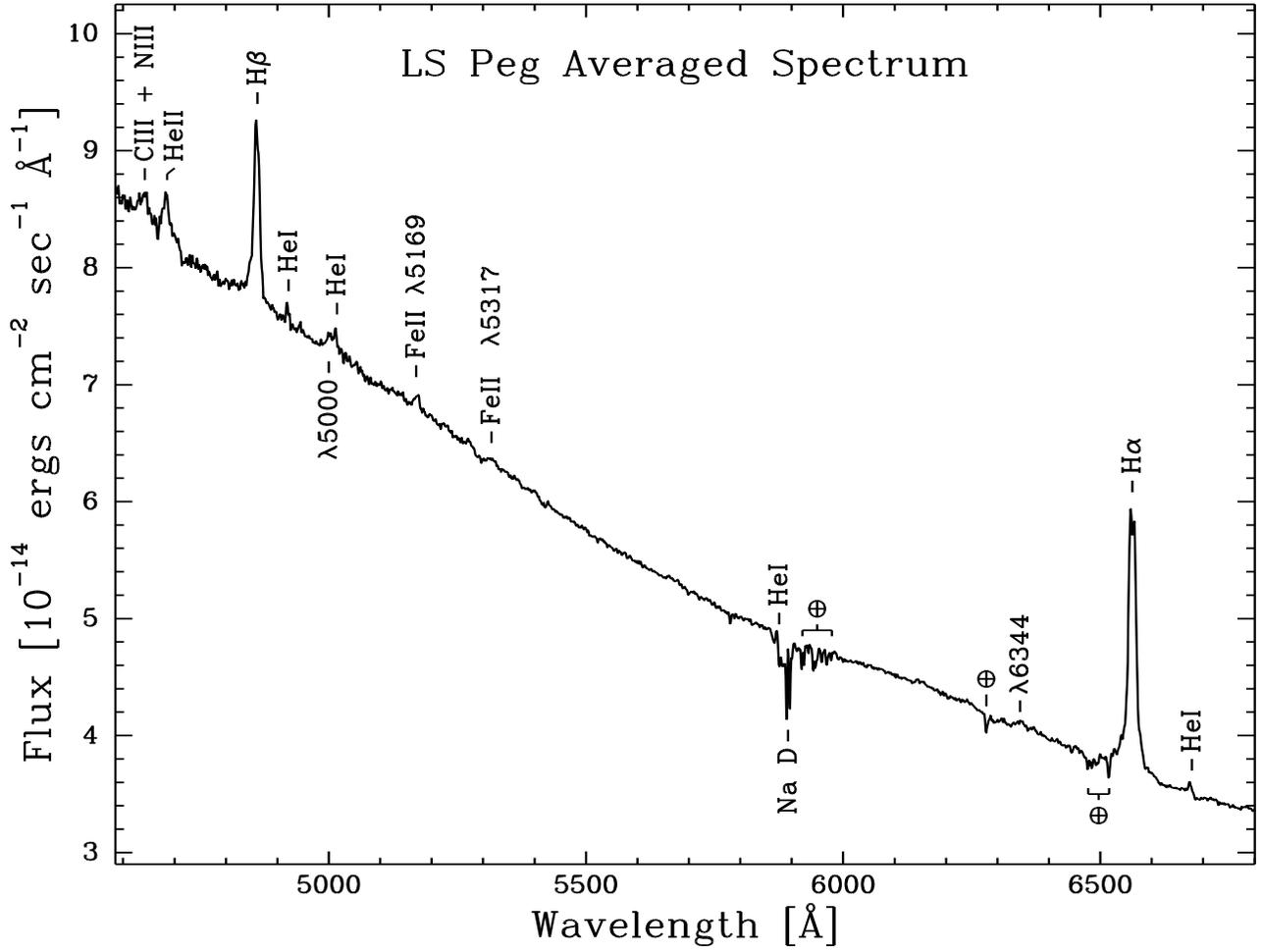}{4.0truein}{270}{70}{70}{-270}{420}
\caption{Averaged spectrum of LS Peg obtained in 1996 October.  The telluric
absorption lines have not been removed. \label{fig4}}
\end{figure}

Figure~\protect\ref{fig4} is an average of 23 spectra from the 1996 October 
run.  The spectrum has typical CV lines (see Table~\protect\ref{tbl-3}), 
including the \ion{C}{3} and \ion{N}{3} blend near $\lambda$4640.  There are 
two weak unidentified lines at approximately $\lambda$5000 and $\lambda$6344.  
The line near $\lambda$5000 is most likely a \ion{N}{2} blend.  The feature 
near $\lambda$6344 could be due to \ion{Si}{1} or \ion{Mg}{1}.  Interstellar 
Na D lines are also present. 

\begin{figure}
\plotfiddle{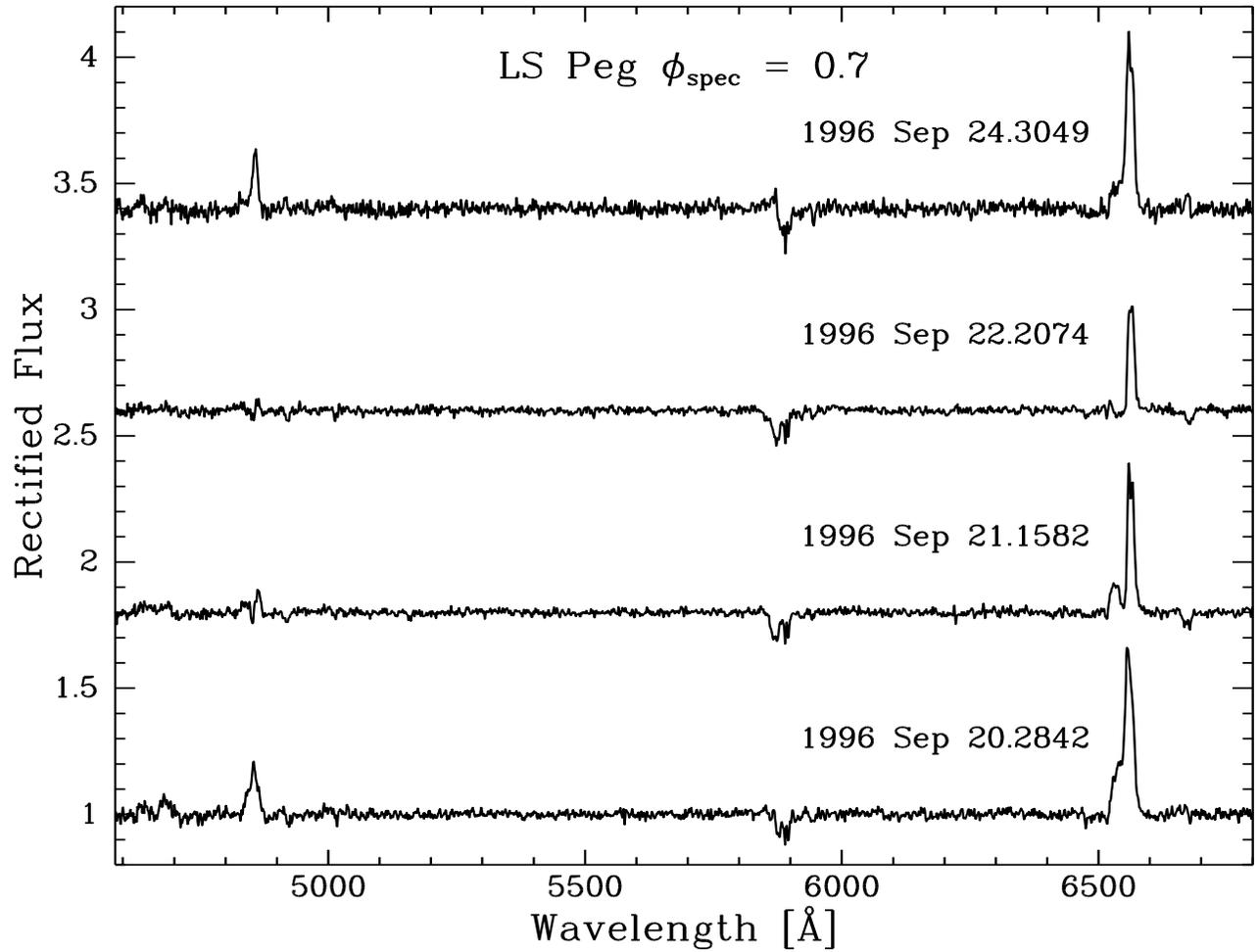}{4.0truein}{270}{70}{70}{-270}{420}
\caption{Rectified spectra of LS Peg from four different orbits at phase
$\phi$ = 0.7.  Top three spectra are offset by a constant flux value. All
four spectra were taken with the same telescope, detector, and chip 
combination. \label{fig5}}
\end{figure}

We spend most of this section discussing phase-dependent features in the line
profiles which reproduce from orbit to orbit, but the profiles also show
secular variations.  Figure~\protect\ref{fig5} shows rectified spectra from 
four different orbits at phase $\phi$ = 0.7.  The middle two spectra show
absorption in the H$\alpha$ and H$\beta$ line wings as well as in the 
\ion{He}{1} $\lambda$6678 and $\lambda$4920 lines, but the upper and lower
spectra show little or no absorption.  In our data, the most pronounced 
orbit-to-orbit variation occurs roughly between phases $\phi_{\rm spec} 
\sim 0.65 - 0.85$, with slight variations mainly in the \ion{He}{1} lines at 
other phases.  These line profile variations are evidently due to the variable 
phasing of the appearance and disappearance of the absorption features. 
Szkody \& Pich\'{e} (~\markcite{sp90}1990) have seen these variations in the 
SW Sex stars DW UMa, SW Sex, and V1315 Aql.

Figures~\protect\ref{fig6}-\ref{fig8} show single-trailed 
spectra of LS Peg.
The single-trailed image was constructed by taking 155 continuum-subtracted 
spectra from the 1996 September and October runs, computing phases from the 
ephemeris above, and binning the spectra into 150 phase bins, each of which 
corresponds to a horizontal line in the image.  To compute the spectrum in 
each phase bin, spectra falling near the central phase of the bin were 
combined using weights computed from a Gaussian in phase, with $\sigma$ = 
0.02 cycle.  Thus a feature appearing in one input spectrum will be smeared 
over several lines of the image. The image lines were stacked to create the 
2-D trailed spectrum with the first 50 lines repeated to provide continuity.  
Note that since there is no eclipse by the secondary, the phase is based on 
the blue-to-red crossing of the broad wings, as described above.

In Figure~\protect\ref{fig6} the broad emission wings in the Balmer lines are
quite prominent.  The width of the emission wings remains approximately 
constant.  The \ion{He}{1} lines all show braided emission and absorption
features, with the absorption crossing over at phase 0.65--0.9, suggestive of
SW Sex stars. The \ion{Fe}{2} $\lambda 5169$ line shows a similar braided 
feature. There also seems to be a faint, broad, high-velocity ($\sim$ 1500
km s$^{-1}$), blueshifted emission feature in the \ion{He}{1} lines starting at
phase 0.4 and ending around phase 1.05; it is most prominent in $\lambda 5876$ 
(see Figure~\protect\ref{fig7}), and is barely visible in $\lambda 6678$, 
$\lambda 5015$, and $\lambda 4920$.  Unfortunately, the interstellar Na D lines
obscure any possible corresponding redshifted emission of $\lambda 5876$, though
close examination of the other lines suggests that there is no redshifted 
emission.  Also, the \ion{He}{1} absorption crosses from red to blue at phase
$\approx$ 0.8.  Both the \ion{C}{3}/\ion{N}{3} blend and \ion{He}{2} lines 
show no phase-dependent features. The constant velocity and sharpness of the 
Na D lines confirms that they are interstellar. 

Figure~\protect\ref{fig8} is a close--up of the H$\alpha$ and \ion{He}{1}
$\lambda 6678$ region of the single trailed spectrum.  The H$\alpha$ wings
extend out to $\pm$2000 km s$^{-1}$ and have a width of approximately 3000 km
s$^{-1}$.  At phases $\phi \approx 0.75 - 0.85$, there is relatively strong 
blueshifted absorption in the H$\alpha$ wing and the edge of the line core.  
This occurs at the same phase as the \ion{He}{1} absorption red-to-blue 
crossing, 
most strongly seen in Figure~\protect\ref{fig7}.  There appears to be fainter 
redshifted absorption in the H$\alpha$ wing and line core at phase $\phi 
\approx 0.25$.  The faint, broad, high-velocity, blueshifted emission in 
$\lambda 6678$ occurs at the same phase as the blueshifted H$\alpha$ wing.  
Note that the maximum blueshift of the faint emission in $\lambda 6678$ occurs
at the same phase as absorption crossing from red to blue (see 
Figure~\protect\ref{fig7} for a clearer example in $\lambda 5876$).  

\section{Doppler Tomography}
We created Doppler tomograms of H$\alpha$ (Figure~\protect\ref{fig9}) and 
\ion{He}{1} $\lambda 6678$ (Figure~\protect\ref{fig10})
using the Fourier-filtered back-projection algorithm (Marsh \& Horne
\markcite{mh88}1988, Horne \markcite{horne91}1991).  Unlike the Maximum Entropy
Method (MEM) algorithm, the Fourier-filtered back-projection algorithm does 
not require only positive values in the single-trailed spectrum
(see the appendix of Harlaftis \& Marsh \markcite{hm96}1996 for a comparison
of the two algorithms). 

We subtracted unity from our single-trailed spectrum (which was created from
continuum-divided spectra) to obtain a mean of zero, as required by the
Doppler tomogram algorithm. The Doppler tomogram is not very sensitive to the 
assumed system parameters; we took $M_{1}= 0.7 M_{\sun}$, $M_{2} = 0.39 
M_{\sun}$, $K_{1} = 120$ km s$^{-1}$, $K_{2} = 215$ km s$^{-1}$, and $i = 
30\arcdeg$. $M_{1}$ and $K_{1}$ are plausible values for SW Sex stars, $M_{2}$ 
is determined from the mass-period relationship ($M_{2} = 
0.065P_{\rm orb}^{5/4}$(h); Warner \markcite{w95}1995), and $i$ is determined 
from our model described in \S5. The model was also used to estimate the 
orbital phasing, so the tomograms could be rotated slightly.

The difference between the H$\alpha$ core and wing tomograms is shown in 
Figure~\protect\ref{fig9}.  The core tomogram (lefthand panel) is similar
to V795 Her (Casares et al. \markcite{c96}1996; Dickinson et al. 
\markcite{d97}1997), PX And (Hellier \& Robinson \markcite{hr94}1994; Still,
Dhillon, \& Jones \markcite{sdj95}1995), and V1315 Aql (Dhillon, Marsh, \& 
Jones \markcite{dmj91}1991; Hellier \markcite{h96}1996).  It suggests that the
line core emission is from the disk edge.  The wing tomogram
(righthand panel) is most similar to V795 Her wing tomogram (Figure 13b in
Dickinson et al. \markcite{d97}1997), and to a lesser extent, SW Sex (Dhillon, 
Marsh, \& Jones \markcite{dmj97}1997) and V1315 Aql (Hellier 
\markcite{h96}1996).  The high-velocity wing emission appears to originate 
from both the stream and the inner disk suggesting the stream-disk interaction
at the re-impact site in the disk-overflow models.  However, the exact stream 
trajectory is uncertain since we don't know $K_{1}$.  The circular pattern is 
an artifact of the phase-smeared noise in the single-trailed spectra. 
The predicted single-trailed spectra reasonably match the data, which suggests 
the emission is near the plane of the disk. 

In our \ion{He}{1} $\lambda6678$ tomogram (Figure~\protect\ref{fig10}), and
to a lesser extent in the H$\alpha$ tomograms, we violate the basic premise of 
Doppler tomography: that the emission/absorption is seen at all phases.  This 
violation is clearly seen in Figure~\protect\ref{fig10} in the predicted trailed
spectrum, in which the absorption extends in phase to produce more of an ``S''
wave not seen in the original data. The emission is similar to V795 Her 
$\lambda6678$ tomograms (Casares et al. \markcite{c96}1996; Haswell et al. 
\markcite{hhtpt94}1994).  The absolute phasing is not known in LS Peg or V795 
Her, so again some rotation of the tomograms may occur.

\section{Model}
In an effort to understand the origin of the very high-velocity Balmer-line
wings, we used a modified version of the disk-overflow model of Hellier and
Robinson \markcite{hr94}(1994). Our purpose was not to exhaustively explore
parameter space, nor to consider radiative transfer problems, but simply to
arrive at a plausible kinematic model. The stream motions were approximated by
particle trajectories, which were computed in a Roche potential using code
described by Thorstensen et al. \markcite{tr91}(1991).  The stellar masses
were held fixed at plausible values ($M_{1}= 0.7 M_{\sun}$, $M_{2} = 0.39 
M_{\sun}$).  In Hellier \& Robinson's scenario, a portion of the accretion 
stream flows over the top of the disk and strikes the disk close to the white 
dwarf. At the point where the stream re-impacts the disk, we assumed the 
emission to be a broad Gaussian with a $\sigma$ equal to 50\% of the average 
of the Keplerian disk and the stream free-fall velocities. As in Hellier and 
Robinson's model, our re-impact region is slightly upstream from the re-impact 
point interpolated from Table 1 in Lubow \markcite{l89}(1989) for our adopted 
parameters. We also assume that prior to re-impact the stream creates 
absorption, which we model as a narrower Gaussian with $\sigma$ equal to 25\% 
of the free-fall velocity.  We then added a double-peaked Keplerian disk 
profile.  Note that in the model, zero phase corresponds to inferior 
conjunction of the secondary, while the phase of the fit to the broad line 
wing motion in the data (hereafter the ``data phase'') has no clear dynamical
interpretation.  Our simulations bore a reasonable resemblence to the data 
using $i = 30{\arcdeg}$, disk radius = $3.5 \times 10^{10}$ cm, and stream 
re-impact region $6.33 - 6.05 \times 10^{9}$ cm from the white dwarf.

Figures~\protect\ref{fig11} and~\protect\ref{fig12} show comparisons of the 
single-trailed H$\alpha$ spectrograms and model.  The model phase was matched
by eye to the data phase.  The model matches the gross features we observe, in 
particular the very high velocity broad wings of the Balmer lines.  It does 
not explain, however, the relatively strong absorption in the wing and edge of 
the line core at data phase $\approx$ 0.8 (model phase $\approx$ 1.4) and to a 
lesser extent at data phase $\approx$ 0.25 (model phase $\approx$ 0.85). Also, 
in the model the line core shows an orbital sinusoidal velocity variation, 
while the observed line core remains at nearly constant velocity (see 
Figure~\protect\ref{fig12}). The absorption in phase with the emission wings 
suggests some absorption associated with the re-impact point.  However when we 
extended the absorption along the entire stream to include the re-impact region
as in the model of V795 Her by Dickinson et al. \markcite{d97}(1997), we 
get absorption in the broad wings at later phases than that is seen in the 
data. But it is important to note that our model is relatively simple.

How does this system compare to others in which large-amplitude 
motion and absorption effects occur?
The effects seen in the SW Sex stars are referred
in phase to the eclipse (SW Sex stars as defined by Thorstensen 
et al. \markcite{tr91}(1991) always show eclipses).
Because LS Peg does not eclipse, we do not know the
absolute orbital phase.  However, our model is computed
using mass points to generate the potential, so the orbital phase
in the model is always known to arbitrary accuracy.  Our
model is matched by eye to the data, giving a rough
relationship between orbital phase and spectroscopic phase.
To the extent our model is realistic, we can ask
whether the line wing motion might be offset in phase
from the `expected' motion of the white dwarf center of mass,
and whether the \ion{He}{1} absorption events occur opposite the
expected `eclipse' phase.  Both these effects are seen
in SW Sex stars (Thorstensen et al. \markcite{tr91}1991).

Using the by-eye match between the model and
our spectroscopic `data phase' (for which
$\phi = 0$ at blue-to-red crossing),
the inferior conjunction of the red star (which would
yield an eclipse in a high-inclination system)
occurs at data phase 0.4.
If the line base traced the
white dwarf motion (unlikely in this case, but often assumed),
the eclipse would be expected at data
phase 0.5.  Thus the spectroscopic
phase `lags' the phase expected for white-dwarf motion by
$\sim 0.1$ cycle.  The sense of the offset is the same as seen
in SW Sex stars, but it is somewhat smaller than the offsets
seen in most SW Sex stars, in which offsets of 0.15 to 0.2
cycles are common (Garnavich et al. \markcite{g90}1990; Honeycutt et al.
\markcite{hsk86}1986; Thorstensen et al. \markcite{tr91}1991;
Casares et al. \markcite{c96}1996).  Again using the model-to-data
phase offset, we find that \ion{He}{1} absorption occurring opposite
the eclipse would appear
around data phase $\phi = 0.9$.  The absorption
is actually seen near data phase $\phi = 0.8$.   In view of the uncertain
correspondence between binary phase and data phase, this is
fair agreement; alternatively, the match to standard SW Sex
behavior could be improved for both the line motion and the
absorption if a slightly different phase offset were assumed.

The model suggests, then, that the relationship
between dynamical (eclipse) phase and
spectroscopic phase is similar to that found in
SW Sex stars.  The absorption seen in the \ion{He}{1} lines
also superficially resembles that seen in SW Sex stars.
To explore how accurately the absorption in LS Peg
resembles the SW Sex phenomenon, we constructed
Figure~\protect\ref{fig13}, which compares the LS Peg spectrogram to a
single-trailed spectrogram of PX And, which was
constructed from data obtained
1988 December and 1989 February (already discussed
by Thorstensen et al. \markcite{tr91}1991).  The PX And data
are of lower spectral resolution than the present
LS Peg data, but the presentation is otherwise
similar.  The resemblance between behaviors
of the \ion{He}{1} $\lambda$5015 absorption in these two stars
is detailed and striking.  Note especially in
PX And how the weak redshifted absorption appears
soon after eclipse, grows in
strength, and moves toward the blue,
with maximum strength occurring opposite the
eclipse. The absorption in LS Peg mimics this
behavior perfectly.

Note that $\phi_{\rm spec}$ is plotted on the
left axes of both panels; these phases are derived
from fits to the H$\alpha$ velocities, and the
two panels are aligned with respect to this phase.
The eclipse phase plotted for PX And is derived from
the observed eclipse, whereas (again) the LS
Peg eclipse phase is {\it inferred} from the
dynamical model of the emission lines.  Again,
there is fair agreement between these two left
axes -- they differ by 0.11 cycle.  Table~\protect\ref{tbl-4} gives
the estimated phases of the \ion{He}{1} absorption events in
both stars in both phase conventions.

In an attempt to constrain the absorption phases, Hellier \markcite{h98}(1998) 
adds a flared disk to the disk-overflow model and, unlike the previous 
versions of the model, includes
spatial obscuration of the disk and stream by the secondary,
obscuration of parts of the disk by the stream, obscuration of the stream by
a flared disk, and obscuration of the stream by other other regions of the
stream.  Previous versions of the model (Hellier \& Robinson 
\markcite{hr94} 1994; Hellier \markcite{h96} 1996) produced the absorption 
features at all phases. In Hellier's recent model, the strength and phase
dependence of the absorption depends on the departure of $i + \alpha$ from 
$90\arcdeg$, where {\it i} is the inclination and $\alpha$ is the flare angle
of the disk.  He found if $i + \alpha \lesssim 80\arcdeg$, the phase-dependent
absorption features weaken and become visible at all phases.  Based on 
theoretical and observational estimates, Hellier assumes a flare angle, 
$\alpha$ $\approx 4\arcdeg$.

We can estimate an upper limit for the inclination of LS Peg. The firmest limit
comes from the absence of eclipses.  Because we don't know
the exact masses, we use the values assumed in our model, $M_{1}= 0.7 M_{\sun}$
and $M_{2} = 0.39 M_{\sun}$, and estimate from Table 1 in Bailey \markcite{ba90}
(1990) that $i \approx 72\arcdeg$, which corresponds to a grazing eclipse.  
For Hellier's model to work at this inclination, the flare angle must be 
greater than $8\arcdeg$, at least twice as large as
the theoretical and observational estimates.  In fact, the inclination of LS 
Peg is probably much lower than $72\arcdeg$ since there is no photometric signal
at the orbital period.  Our model provides a reasonable
fit at an inclination of $30\arcdeg$, which would imply an unrealistic flare
angle of over $50\arcdeg$.

\section{Conclusion and Discussion}
We obtain a radial velocity period of $0.174774 \pm 0.000003$ d.  This is 
almost certainly the orbital period.

The phase-dependent absorption is reminiscent of SW Sex stars, and a comparison
between our data, our model, and archival data of the well-studied SW Sex star
PX And suggests that the absolute binary phase in LS Peg is about as expected
for the SW Sex phenomenon.  There is a striking qualitative resemblance 
between the phase-dependent \ion{He}{1} absorption features observed in LS
Peg and those observed in SW Sex stars.

As new objects are discovered, the SW Sex behaviors
listed by Thorstensen et al. \markcite{tr91}(1991) continue to appear together
as a suite.   Although
eclipses are formally necessary in the original
definition of SW Sex stars, phase-dependent
\ion{He}{1} absorption appears to be fairly
common in non-eclipsing
bright novalikes (such as WX Ari; Beuermann et al. \markcite{b92}1992).
The formal criterion that a system be eclipsing
for admission to the SW Sex class was always
somewhat unsatisfactory, as it depends on our
accidental viewing angle rather than on a physical
characteristic of the system; in this sense
the phase-dependent
absorption features are a more fundamental
criterion.  We believe that LS Peg should be
classified as an SW Sex star even though it does not
eclipse.

\begin{figure}
\plotfiddle{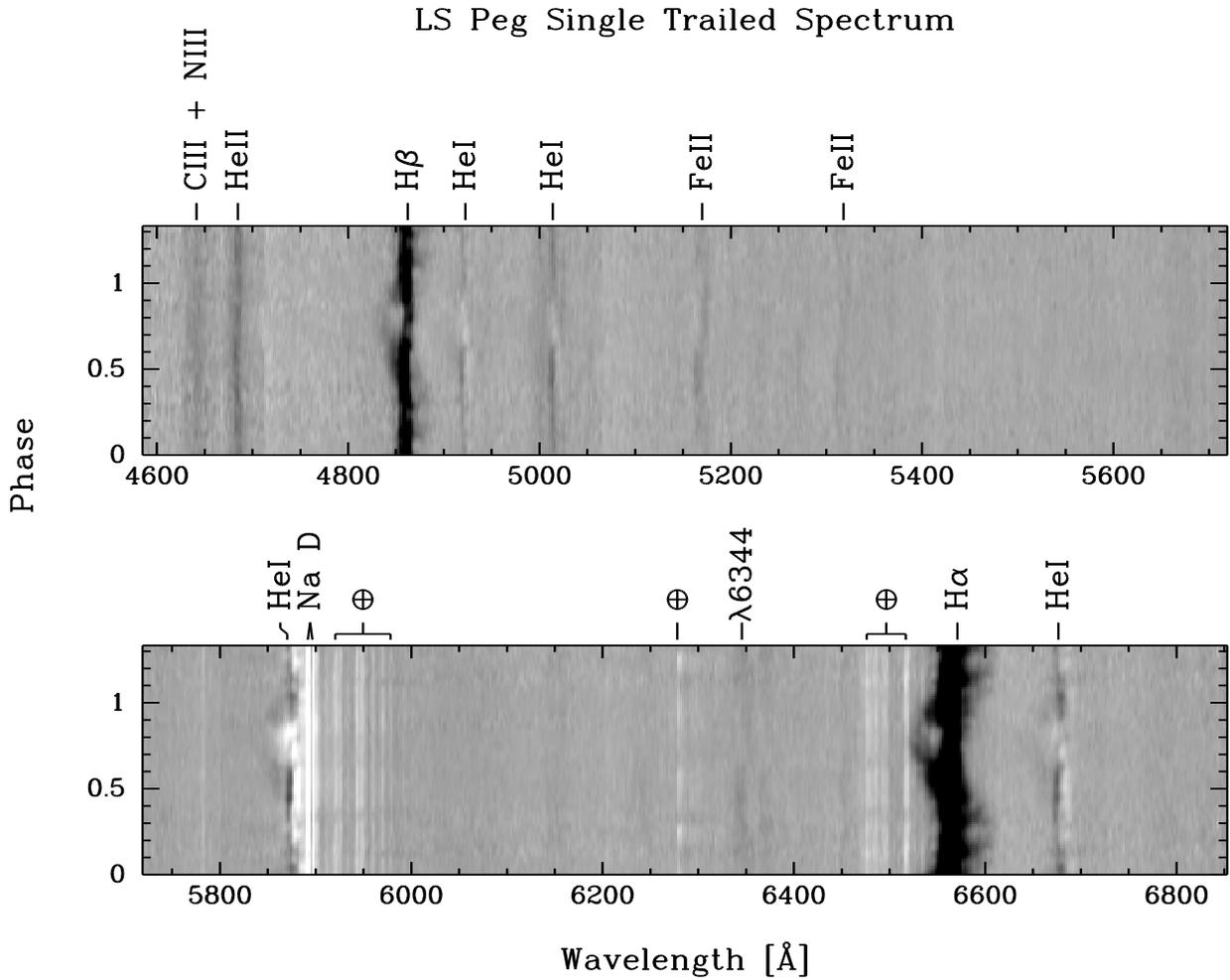}{4.0truein}{270}{70}{70}{-270}{420}
\caption{Single-trailed spectrum of LS Peg. One hundred fifty-five $\sim$ 5
min exposures from 1996 Sept. and Oct. are folded into 150 phase bins with
a Gaussian phase smearing of $\sigma$ = 0.02 cycles.  The first 50 bins
are repeated for continuity. This is a negative image (black = emission). The
telluric absorption lines have not been removed. \label{fig6}}
\end{figure}

\begin{figure}
\plotfiddle{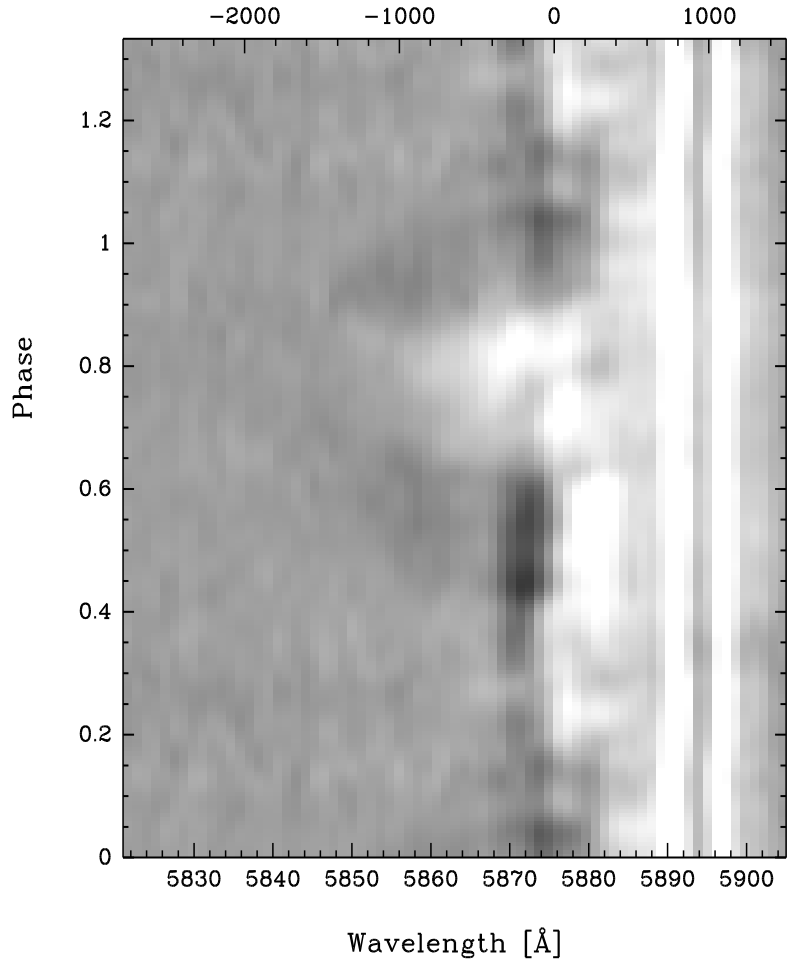}{4.0truein}{0}{60}{60}{-180}{-20}
\caption{Detail of the \protect\ion{He}{1} $\lambda$5876 region of the
single-trail spectrum.  Bottom scale is wavelength in {\AA} and top scale is
$\lambda$5876 radial velocity in km s$^{-1}$.  The interstellar Na D lines are 
prominent. The broad, high-velocity, blueshifted emission occurs at the same 
phase as the blueshifted wings of the Balmer lines.  \label{fig7}}
\end{figure}

\begin{figure}
\plotfiddle{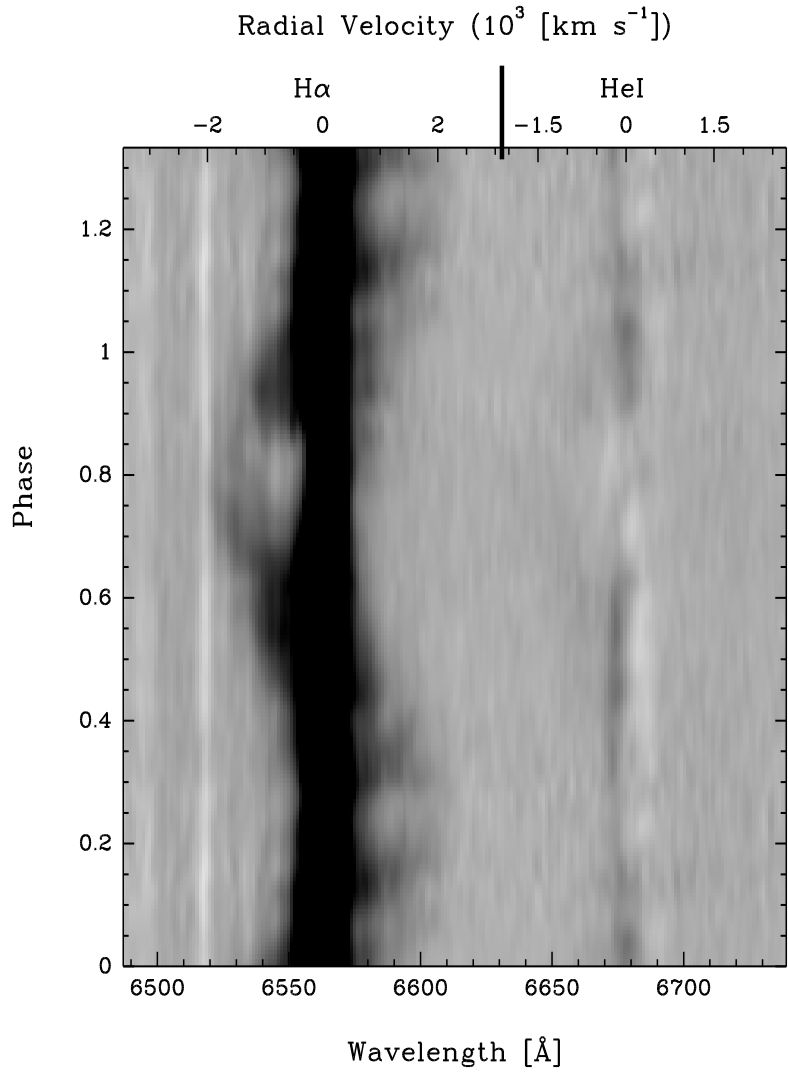}{4.0truein}{0}{60}{60}{-195}{-20}
\caption{Detail of the H$\alpha$ and \protect\ion{He}{1} $\lambda$6678 
region of the single-trail spectrum.  Bottom scale is wavelength in {\AA} and 
top scale is radial velocity in $10^{3}$ km s$^{-1}$. The solid line separates 
the radial velocity scales for H$\alpha$ and \protect\ion{He}{1}.  Note the 
terrestial absorption line blueward of H$\alpha$. \label{fig8}}
\end{figure}

\begin{figure}
\plotfiddle{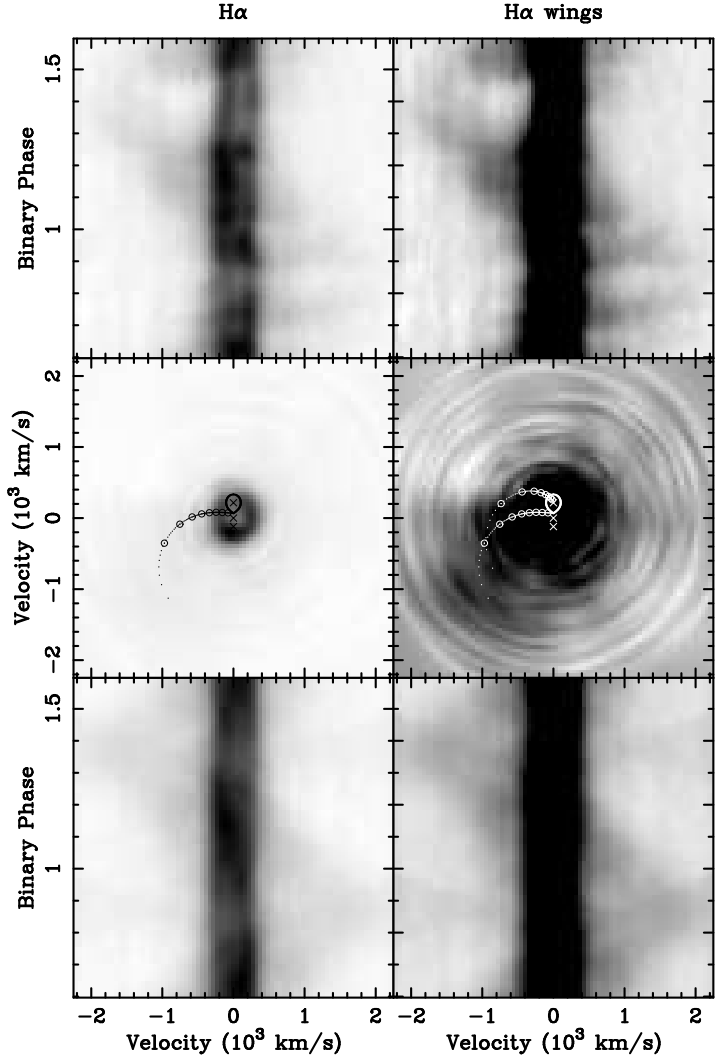}{4.0truein}{0}{55}{55}{-170}{-20}
\caption{H$\alpha$ single-trailed spectra {\it(top)}, Doppler 
tomograms {\it(middle)}, and predicted single-trailed spectra from the 
tomograms {\it(bottom)}.  The phasing of the spectra run from 0.6 to 1.6. The 
intensity in these negative images (black = emission) is adjusted in the left 
panels to emphasize the line core, while the right panels emphasize the line 
wings.  Crosses mark the center of mass of the secondary {\it(top)}, the system
{\it(middle)}, and the white dwarf {\it(bottom)}. The secondary Roche lobe is
outlined.  On the left, the gas stream trajectory is plotted, while on the 
right, the upper trajectory is the Keplerian velocity of the disk along the 
gas stream and the lower trajectory is the gas stream. \label{fig9}}
\end{figure}

\begin{figure}
\plotfiddle{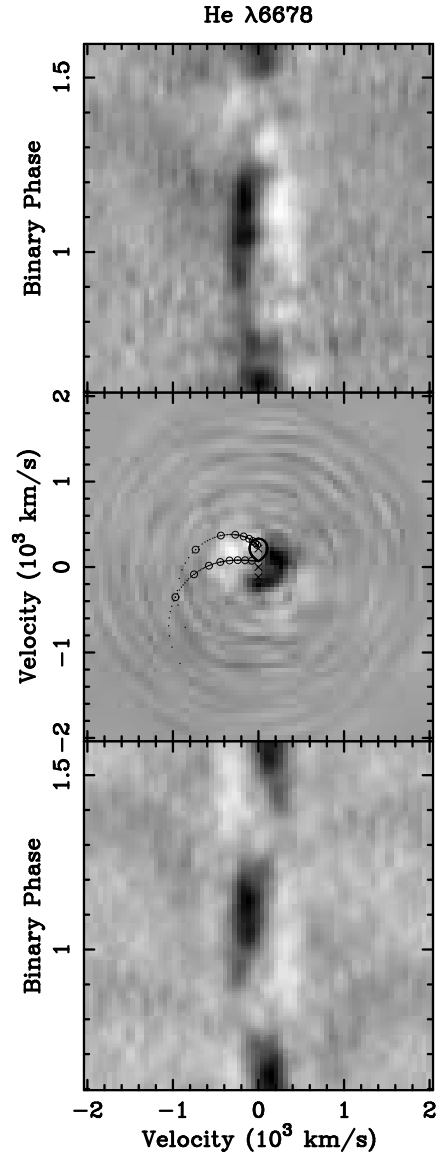}{4.0truein}{0}{60}{60}{-180}{-20}
\caption{\protect\ion{He}{1} $\lambda6678$ single-trailed spectrum, Doppler
tomogram, and predicted single-trailed spectrum with same conventions 
as Figure~\protect\ref{fig9}.  \label{fig10}}
\end{figure}

\begin{figure}
\plotfiddle{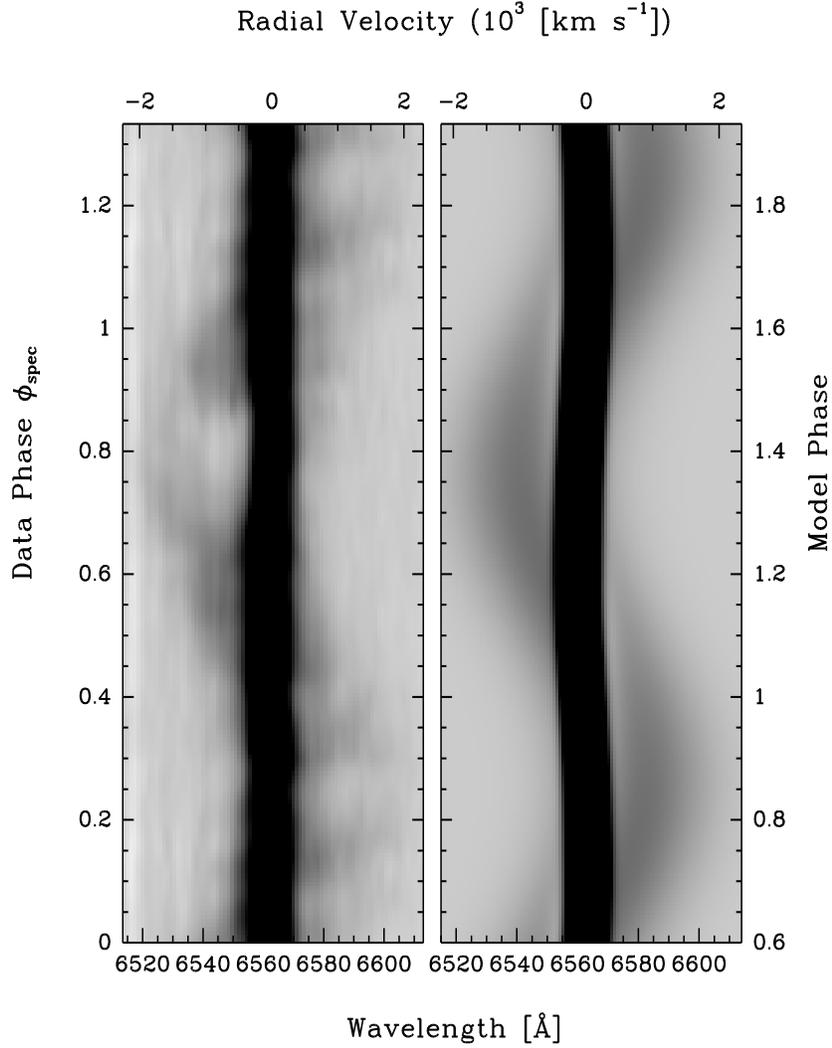}{4.0truein}{0}{60}{60}{-185}{-20}
\caption{Comparison of the single-trailed H$\alpha$ spectrogram {\it(left)} 
with the disk-overflow model {\it(right)}.  The zero phase of the 
``data phase'' is the blue-to-red crossing of the H$\alpha$ wings, while the 
zero phase of the model corresponds to the inferior conjunction of the 
secondary. Intensity is adjusted to emphasize the high-velocity wings.  
\label{fig11}}
\end{figure}

\begin{figure}
\plotfiddle{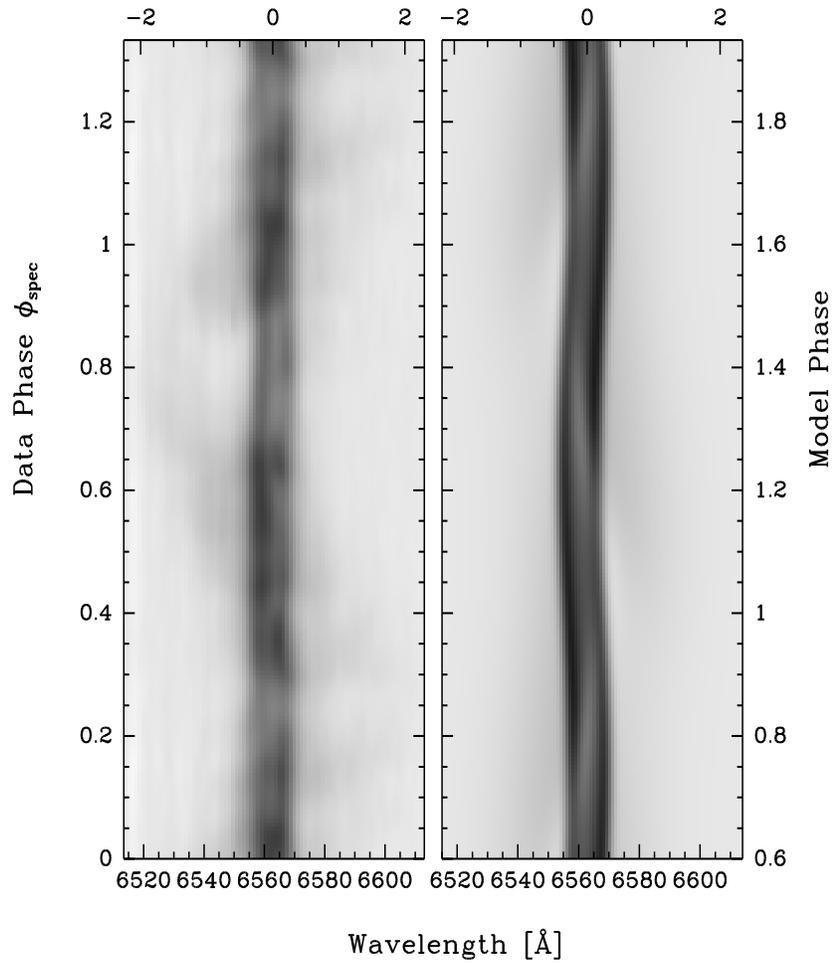}{4.0truein}{0}{60}{60}{-185}{-20}
\caption{Same as Figure~\protect\ref{fig11}, but with the intensity adjusted 
to emphasize the line cores.  \label{fig12}}
\end{figure}

\begin{figure}
\plotfiddle{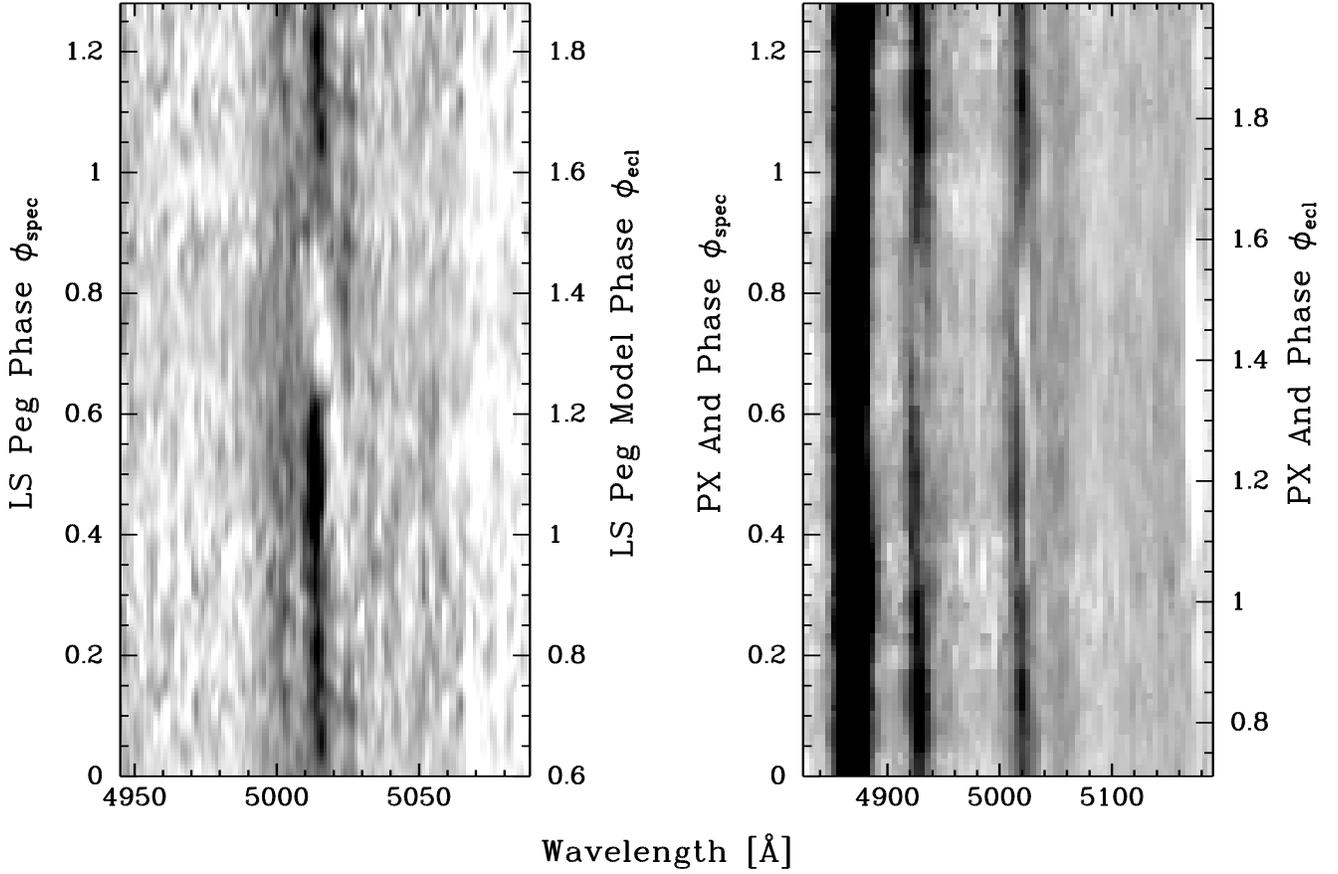}{4.0truein}{270}{70}{70}{-275}{400}
\caption{Comparison of LS Peg and PX And single-trailed spectra in the
$\lambda$5015 region.  Phase zero for $\phi_{\rm spec}$ corresponds to the 
blue-to-red crossing of H$\alpha$ and phase zero for $\phi_{\rm ecl}$ 
corresponds to the eclipse.  The PX And data were previously published by 
Thorstensen et al.~\protect\markcite{tr91}(1991) and are folded into 
100 phase bins with a Gaussian phase smearing of $\sigma$ = 0.02 cycles. 
\label{fig13}}
\end{figure}

The (modest) success of our simple kinematic model
in accounting for the broad line wings -- and the
indirect evidence that the absolute binary
phase predicted by the model is as expected
given the behavior of other SW Sex stars --
suggests strongly that the disk-overflow model
proposed by Hellier \& Robinson \markcite{hr94}(1994) is essentially correct.
The striking broad line wings seen in Figures~\protect\ref{fig6},
~\protect\ref{fig8}, and ~\protect\ref{fig11}
are almost certainly coming from material associated
with the stream overflowing the disk and the
stream re-impact point.
 
However, as Casares et al. \markcite{c96}(1996) pointed out, Hellier \&
Robinson's disk-overflowing stream scenario does not limit the phase of the
absorption to around $\phi \sim 0.5$.  The inclination of LS Peg is so low
that a flared disk such as that proposed by Hellier \markcite{h98}(1998)
{\it cannot} explain the phase dependence of the absorption, so a new
explanation must be sought.  A clue may come from recent 
hydrodynamical simulations of the stream-disk interaction by Armitage \&
Livio (\markcite{al96}1996, \markcite{al98}1998).  For inefficent cooling, an 
explosive stream-disk interaction produces a ``halo'' of material above and 
below the disk as well as a bulge along the disk rim downstream of the stream 
impact.  Because of the surface brightness of the disk is highest toward its
center, the absorption produced by such out-of-plane material could be highly
dependent on viewing angle.

We see no evidence of a strictly periodic 
modulation in the photometry.  In view of the lack of polarization, the lack 
of a strict photometric period, and the similarity to SW Sex stars (none of
which are known to be magnetic), there is little to corroborate suggestions
that this is a magnetic CV.
Thus we can consider it unlikely that the magnetic accretion models of Casares 
et al. \markcite{c96}(1996) and Williams \markcite{w89}(1989) apply in this 
case.

\acknowledgements
We warmly thank Bob Fried at Braeside Observatory and the Columbia
University Astronomy C3997 Fall 1995 class for taking the photometric data.
We also thank Keith Horne 
for providing the Fourier-filtered back-projection Doppler tomography code.  
We thank the NSF for support through grant AST-9314787
and the MDM staff for their usual excellent support.  
This research made use of the Simbad database, operated at CDS, Strasbourg, 
France.

\clearpage

\begin{deluxetable}{clccccc}
\small
\tablecaption{Observation Log \label{tbl-1}}
\tablewidth{0pt}
\tablehead{
\colhead{Dates (UT)} & \colhead{No. of} & \colhead{Telescope} &
\colhead{Spectr.} & \colhead{CCD} & \colhead{$\lambda$ range} & 
\colhead{FWHM} \\
\colhead{} & \colhead{Spectra} & \colhead{(m)} & \colhead{} & \colhead{} &
\colhead{(\AA)} & \colhead{(\AA)}
}
\startdata
1996 Jul. & 19 & 2.4 & modular & Tek $1024^{2}$ thinned & 4610 - 6688 & 3 \nl
1996 Sep. & 71 & 1.3 & MkIII & Loral $2048^{2}$ unthinned & 4310 - 7184 & 5 \nl
1996 Oct. & 68 & 2.4 & modular & Loral $2048^{2}$ unthinned & 4585 - 7171 & 
3 \nl
1996 Dec. & 17 & 1.3 & MkIII & Tek $1024^{2}$ thinned & 4472 - 6760 & 5 \nl
1997 Jun. & 4  & 2.4 & modular & Tek $2048^{2}$ thinned & 4000 - 7502 & 3.5 \nl
\enddata
\end{deluxetable}

\clearpage

\begin{deluxetable}{lrlrlrlr}
\small
\tablewidth{0pt}
\tablecolumns{8}
\tablecaption{LS Peg H$\alpha$ Line Based Radial Velocities \label{tbl-2}}
\tablehead{
\colhead{HJD\tablenotemark{a}} & \colhead{V} & \colhead{HJD\tablenotemark{a}}
& \colhead{V} & \colhead{HJD\tablenotemark{a}} & \colhead{V} &
\colhead{HJD\tablenotemark{a}} & \colhead {V} \\
\colhead{} &\colhead{(km s$^{-1}$)} & \colhead{} & \colhead{(km s$^{-1}$)} &
\colhead{} &\colhead{(km s$^{-1}$)} & \colhead{} & \colhead{(km s$^{-1}$)}
}
\startdata
 $266.9369$ &  $  -770$   &  $347.6384$ &  $  -354$   &  $353.8180$ &  $  -486$   &  $358.8963$ &  $     3$   \nl 
 $266.9413$ &  $  -738$   &  $347.6441$ &  $  -761$   &  $353.8220$ &  $  -385$   &  $359.7101$ &  $  -477$   \nl 
 $266.9463$ &  $  -565$   &  $347.6525$ &  $  -555$   &  $353.8260$ &  $  -152$   &  $359.7127$ &  $  -796$   \nl 
 $267.9184$ &  $   125$   &  $347.6582$ &  $  -615$   &  $353.8300$ &  $   -79$   &  $359.7153$ &  $  -609$   \nl 
 $267.9248$ &  $   -21$   &  $347.6639$ &  $  -879$   &  $353.8339$ &  $   473$   &  $359.7643$ &  $  -331$   \nl 
 $267.9292$ &  $  -117$   &  $347.6696$ &  $  -637$   &  $353.8379$ &  $   583$   &  $359.7690$ &  $  -308$   \nl 
 $267.9404$ &  $  -409$   &  $347.7694$ &  $   642$   &  $353.8419$ &  $   400$   &  $359.7737$ &  $  -125$   \nl 
 $267.9458$ &  $  -670$   &  $347.7751$ &  $   404$   &  $354.8731$ &  $  -221$   &  $359.7783$ &  $   240$   \nl 
 $269.9007$ &  $  -592$   &  $347.7808$ &  $   336$   &  $354.8771$ &  $  -253$   &  $359.8520$ &  $   222$   \nl 
 $269.9044$ &  $  -505$   &  $347.7865$ &  $   -25$   &  $354.8810$ &  $   -20$   &  $359.8546$ &  $   154$   \nl 
 $269.9073$ &  $  -459$   &  $347.7922$ &  $    43$   &  $355.8596$ &  $  -331$   &  $359.8572$ &  $   195$   \nl 
 $270.8751$ &  $   404$   &  $347.7979$ &  $  -112$   &  $355.8635$ &  $  -116$   &  $359.8597$ &  $  -239$   \nl 
 $270.8794$ &  $   248$   &  $347.8061$ &  $  -345$   &  $355.8675$ &  $  -454$   &  $359.8623$ &  $  -152$   \nl 
 $270.8999$ &  $  -208$   &  $347.8118$ &  $  -674$   &  $356.8681$ &  $   496$   &  $359.8649$ &  $  -239$   \nl 
 $270.9042$ &  $  -281$   &  $347.8175$ &  $  -715$   &  $356.8721$ &  $   144$   &  $359.8675$ &  $  -426$   \nl 
 $270.9122$ &  $  -427$   &  $347.8232$ &  $  -432$   &  $356.8761$ &  $   341$   &  $359.8701$ &  $  -262$   \nl 
 $270.9179$ &  $  -651$   &  $347.8289$ &  $  -491$   &  $356.8800$ &  $   163$   &  $359.8727$ &  $  -394$   \nl 
 $270.9772$ &  $  -446$   &  $347.8347$ &  $  -665$   &  $356.8840$ &  $   108$   &  $359.8753$ &  $  -504$   \nl 
 $270.9816$ &  $  -263$   &  $347.8423$ &  $  -615$   &  $356.8880$ &  $   -57$   &  $360.5978$ &  $  -650$   \nl 
 $346.7405$ &  $   413$   &  $347.8480$ &  $  -710$   &  $356.8919$ &  $  -253$   &  $360.6018$ &  $  -806$   \nl 
 $346.7462$ &  $   199$   &  $347.8537$ &  $  -519$   &  $356.8959$ &  $  -349$   &  $360.6057$ &  $  -737$   \nl 
 $346.7519$ &  $  -158$   &  $347.8595$ &  $  -432$   &  $356.8999$ &  $  -280$   &  $360.6097$ &  $  -486$   \nl 
 $346.7576$ &  $   121$   &  $347.8652$ &  $  -569$   &  $356.9038$ &  $  -513$   &  $360.6137$ &  $  -550$   \nl 
 $346.7633$ &  $  -190$   &  $347.8709$ &  $  -372$   &  $357.7301$ &  $   629$   &  $434.5914$ &  $   577$   \nl 
 $346.7691$ &  $  -372$   &  $350.8049$ &  $  -829$   &  $357.7353$ &  $   446$   &  $434.5968$ &  $    70$   \nl 
 $346.7785$ &  $  -578$   &  $350.8106$ &  $  -587$   &  $357.7379$ &  $   496$   &  $434.6022$ &  $   476$   \nl 
 $346.7842$ &  $  -482$   &  $350.8171$ &  $  -633$   &  $357.7405$ &  $   350$   &  $435.5937$ &  $  -588$   \nl 
 $346.7899$ &  $  -692$   &  $350.8228$ &  $  -363$   &  $357.7430$ &  $   368$   &  $435.5991$ &  $  -341$   \nl 
 $346.7956$ &  $  -432$   &  $350.8343$ &  $  -414$   &  $357.8829$ &  $   126$   &  $435.6046$ &  $  -643$   \nl 
 $346.8013$ &  $  -414$   &  $350.8475$ &  $  -144$   &  $357.8855$ &  $  -253$   &  $435.6100$ &  $  -414$   \nl 
 $346.8070$ &  $  -528$   &  $350.8532$ &  $    -7$   &  $357.8907$ &  $   -79$   &  $438.5739$ &  $  -560$   \nl 
 $346.8156$ &  $  -578$   &  $350.8589$ &  $   -25$   &  $357.8932$ &  $   368$   &  $438.5793$ &  $  -506$   \nl 
 $346.8270$ &  $  -121$   &  $350.8646$ &  $   121$   &  $357.8958$ &  $   186$   &  $438.5847$ &  $  -241$   \nl 
 $346.8327$ &  $   -94$   &  $351.7244$ &  $  -235$   &  $358.8452$ &  $  -417$   &  $438.5901$ &  $   -99$   \nl 
 $346.8384$ &  $    34$   &  $351.7301$ &  $  -149$   &  $358.8484$ &  $  -641$   &  $440.5717$ &  $   499$   \nl 
 $346.8441$ &  $  -130$   &  $351.7358$ &  $    94$   &  $358.8517$ &  $  -678$   &  $440.5771$ &  $   335$   \nl 
 $346.8521$ &  $   -16$   &  $351.7415$ &  $   491$   &  $358.8550$ &  $  -618$   &  $440.5825$ &  $   454$   \nl 
 $346.8578$ &  $    34$   &  $351.7472$ &  $   683$   &  $358.8648$ &  $  -532$   &  $440.5879$ &  $    47$   \nl 
 $346.8635$ &  $   683$   &  $351.7529$ &  $   217$   &  $358.8681$ &  $  -532$   &  $440.5933$ &  $  -158$   \nl 
 $346.8692$ &  $   642$   &  $351.7607$ &  $   683$   &  $358.8747$ &  $  -230$   &  $440.5987$ &  $   326$   \nl 
 $346.8749$ &  $   701$   &  $351.7664$ &  $   578$   &  $358.8797$ &  $  -417$   &  $626.9807$ &  $  -710$   \nl 
 $347.6156$ &  $   258$   &  $351.7721$ &  $   623$   &  $358.8832$ &  $   -79$   &  $627.9839$ &  $  -371$   \nl 
 $347.6213$ &  $   157$   &  $351.7778$ &  $   760$   &  $358.8865$ &  $    -2$   &  $628.9923$ &  $   145$   \nl 
 $347.6270$ &  $  -149$   &  $351.7835$ &  $   569$   &  $358.8898$ &  $    76$   &  $629.9859$ &  $   515$   \nl 
 $347.6327$ &  $  -459$   &  $351.7892$ &  $   569$   &  $358.8930$ &  $    -2$   &  \nl 
\enddata
\tablenotetext{a}{Heliocentric JD of mid-integration minus 2450000.}
\end{deluxetable}

\clearpage

\begin{deluxetable}{ll}
\small
\tablecaption{Equivalent Widths \label{tbl-3}}
\tablewidth{0pt}
\tablecolumns{2}
\tablehead{
\colhead{Lines}  &  \colhead{EW (\AA)}
}
\startdata
H$\alpha$ & 12.6 \nl
H$\beta$ & 2.6 \nl
\ion{He}{1} $\lambda 6678$ & 0.2 \nl
\ion{He}{1} $\lambda 5015$ & 0.2 \nl
\ion{He}{1} $\lambda 4920$ & 0.1 \nl
\ion{He}{2} $\lambda 4686$ & 0.6 \nl
\ion{C}{3}+\ion{N}{3} $\lambda 4640-4650$ & 0.5 \nl
\ion{Fe}{2} $\lambda 5169$ & 0.2 \nl
\ion{Fe}{2} $\lambda 5317$ & 0.1 \nl
$\lambda 5000$\tablenotemark{a} & 0.2 \nl
$\lambda 6344$\tablenotemark{b} & 0.1 \nl
\enddata
\tablenotetext{a}{Possible blend of \ion{N}{2}.}
\tablenotetext{b}{\ion{Si}{1}? or \ion{Mg}{1}?}
\end{deluxetable}

\begin{deluxetable}{lcccc}
\small
\tablecaption{LS Peg and PX And ~\protect\ion{He}{1} $\lambda5015$ Phase 
Measurements \label{tbl-4}}
\tablehead{
\colhead{} & \multicolumn{2}{c}{PX And} & \colhead{Model} & \colhead{LS Peg} \\
\colhead{} & \colhead{$\phi_{\rm ecl}$} & \colhead{$\phi_{\rm spec}$} & 
\colhead{$\phi_{\rm ecl}$} & \colhead{$\phi_{\rm spec}$}   
}
\tablecolumns{5}
\startdata
inferior conj. & 0.0 & 0.29 & 0.0 & 0.40 \nl
start of deeper absn & 0.41 & 0.70 & [0.23]\tablenotemark{a} & 0.63 \nl
center of deeper absn & 0.44 & 0.79 & [0.31]\tablenotemark{a} & 0.71 \nl
\enddata
\tablenotetext{a}{Estimated from Figure~\protect\ref{fig10}.}
\end{deluxetable}

\clearpage

\end{document}